%% file: HIJpsi_PLB.tex
\documentclass[preprint,12pt]{elsarticle}
\newcommand{\PbPb}{\mbox{lead-lead} }

\usepackage{amssymb}
\usepackage{lineno}

\begin{document}

\begin{frontmatter}



\title{Measurement of the centrality dependence of $J/\psi$ yields and observation of $Z$ production in lead-lead collisions with the ATLAS detector at the LHC}

\author{G. Aad \textit{et al.} (The ATLAS Collaboration)}

\author{}



\begin{abstract}
Using the ATLAS detector, a centrality-dependent
suppression has been observed in the yield of $J/\psi$ mesons produced in the collisions of
lead ions at the Large Hadron Collider.  In a sample of minimum-bias lead-lead
collisions at a nucleon-nucleon centre of mass energy $\sqrt{s_{NN}}=$ 2.76~TeV,
corresponding to an integrated luminosity of about $6.7~\mu \mathrm{b}^{-1}$, $J/\psi$
mesons are reconstructed via their decays to $\mu^+\mu^-$ pairs.  
The measured $J/\psi$ yield, normalized to the number of binary nucleon-nucleon collisions, 
is found to significantly decrease from peripheral to central collisions.
The centrality dependence is found to
be qualitatively similar to the trends observed at previous, lower energy experiments.
The same sample is used to reconstruct $Z$ bosons in the $\mu^+\mu^-$ final state, and a
total of 38 candidates are selected in the mass window 
of 66 to 116 GeV.
The relative $Z$ yields as a function of centrality are also presented,
although no conclusion can be inferred 
about their scaling with the number of binary collisions, 
because of limited statistics.
This analysis provides the first results on $J/\psi$ and $Z$ production in
lead-lead collisions at the LHC. 

\end{abstract}

\begin{keyword}
ATLAS \sep
LHC \sep
Heavy Ions \sep
J/psi \sep
Z Boson \sep
Centrality dependence
\end{keyword}

\end{frontmatter}


\section{Introduction}
\label{Introduction}

The measurement of quarkonia production in ultra-relativistic heavy
ion collisions provides a potentially powerful tool for studying the
properties of hot and dense matter created in these collisions.  If
deconfined matter is indeed formed, then
colour screening is expected to prevent the formation of quarkonium
states when the screening length becomes shorter than the
quarkonium size~\cite{Matsui:1986dk}. Since this
length is directly related to the temperature, a measurement of 
a suppressed quarkonium yield 
may provide direct experimental sensitivity to the temperature of the
medium created in high energy nuclear collisions~\cite{Mocsy:2007jz}.

The interpretation of $J/\psi$ suppression in terms of colour screening
is generally complicated by  the quantitative agreement  
between the overall levels of $J/\psi$ suppression measured by the
NA50 experiment at the CERN SPS~\cite{Alessandro:2004ap}
($\sqrt{s_{NN}}=17.3$ GeV) and the
PHENIX experiment at RHIC~\cite{Adare:2006ns} ($\sqrt{s_{NN}}=200$ GeV).
Data from proton-nucleus and deuteron-gold collisions also show 
decreased rates of $J/\psi$ production~\cite{pAJpsi}, indicating
that other mechanisms may come into play.
Finally, there exist proposals for $J/\psi$ {\it enhancement}
at high energies from charm quark recombination~\cite{Thews:2005vj}.
Measurements at higher energies, with concomitantly higher temperatures and
heavy quark production rates, are clearly needed to address these
debates with new experimental input.
The production of $Z$ bosons, only available in heavy ion collisions at
LHC energies, can serve as a reference process for $J/\psi$ production, since $Z$'s are not
expected to be affected by the hot, dense medium, although modifications to
the nuclear parton distribution functions must be considered~\cite{Vogt:2000hp}.

The LHC heavy ion program, which commenced in November 2010, 
offers an opportunity to measure $J/\psi$ and $Z$ production in nuclear
collisions at the highest energies ever achieved. The ATLAS detector
provides excellent muon detection capabilities 
down to momenta 
of about 3~GeV, and $J/\psi$ mesons and $Z$ bosons can be readily detected 
via their decays to $\mu^+ \mu^-$ final states. This Letter presents the
first measurements of
the relative yields of $J/\psi$ meson and $Z$ boson
decays in lead-lead collisions at a
nucleon-nucleon center of mass energy of  $\sqrt{s_{NN} }=
2.76$~TeV. The yields
are measured in four bins of collision centrality, and the
variation of the yields with centrality is compared to the
dependence expected if hard scattering processes scale according to expectations 
from nuclear geometry. 
No attempts are made to account for 
``normal nuclear suppression''~\cite{Alessandro:2004ap}, nor for 
feed-down of $J/\psi$ from higher mass charmonium states or $B$ hadron decay.

\section{Di--muon event selection}
\label{Samples}

Muons are measured by combining independent
measurements of the muon trajectories from the Inner Detector (ID) and  
the Muon Spectrometer (MS). A detailed description of these
detectors and their performance
in proton-proton collisions can
be found in Refs.~\cite{AtlasDetector,AtlasPerformance}.
The ID volume is within the 2~T field of a superconducting solenoid,
and measures the trajectories
of charged particles in the pseudorapidity region $|\eta|<2.5$~\footnote{
In the right-handed ATLAS coordinate system, the pseudorapidity $\eta$ is defined 
as $\eta = - \ln[\tan(\theta/2)]$, where the polar angle $\theta$ is 
measured with respect to the LHC beamline. 
The azimuthal angle $\phi$ is measured with respect to the x-axis, which points 
towards the centre of the LHC ring. The z-axis is parallel to the anti-clockwise 
beam viewed from above.  Transverse momentum and energy are defined 
as $p_\mathrm{T} = p \sin \theta$ and $E_\mathrm{T} = E \sin \theta$, respectively.
}. 
A charged particle
typically traverses three layers of silicon pixel detectors, eight 
silicon strip sensors (SCT detector) arranged in four layers of double-sided
modules, and a transition radiation
tracker composed of straw tubes. The MS surrounds the calorimeters and provides
tracking for muons with $|\eta|<2.7$ and triggering in
the range  $|\eta|<2.4$. 
The muon momentum determination is based on three stations of precision
drift chambers that measure the trajectory of each muon in a toroidal
magnetic field produced by three air-core toroids. 
In order to reach the MS, muons have to cross the electromagnetic and
hadronic calorimeters, losing typically 3 to 5 GeV of energy,
depending on the muon pseudorapidity. The calorimeters efficiently 
absorb the copious charged and neutral hadrons produced in lead-lead 
collisions.

The trigger system has three stages, the first of which (Level-1)
is hardware based.
The Level-1 minimum-bias trigger uses either the two sets of
Minimum-Bias Trigger Scintillator (MBTS) counters,
covering $2.1<|\eta|<3.9$ on each side of the experiment, 
or the two Zero-Degree Calorimeters (ZDC), each
positioned at 140~m from the collision point, detecting
neutrons and photons with $|\eta|>8.3$.  
No muon-specific 
triggers were used to select the data presented here.
The MBTS trigger was configured to require at least
one hit above threshold from each side of the detector. 
A Level-2 
timing requirement on a coincidence of signals from the MBTS
was then imposed to remove beam backgrounds. The trigger efficiency was
studied using an independent trigger probing random filled bunch
crossings at Level-1. 
For these triggers, empty events were removed by testing for a minimal level of
activity in the silicon detectors. The combined trigger and
event selection efficiency is discussed in section~\ref{Rcoll}.

In the offline analysis, minimum-bias triggered events are required to
have a reconstructed primary vertex, at least one hit in each set of
MBTS counters, and a time difference between the sides of less than
$3$~ns to reject beam-halo and other beam-related background
events.  Measurements of the muon trajectories from both the ID and 
MS are combined, resulting in a relative momentum  resolution ranging
from about 2\% at low momentum up to about 3\% at $p_{\mathrm{T}} \sim 50$~GeV.
For this analysis, oppositely charged muons
are selected with a minimum $p_{\mathrm{T}}$ of 3 GeV each and within the region
$|\eta|<2.5$. 

The data sample consists of approximately 6.7~$\mu \mathrm{b}^{-1}$ from the 
2010 LHC heavy ion run.
In order to determine the $J/\psi \to \mu^{+} \mu^{-}$ 
reconstruction efficiency, Monte Carlo (MC) samples have been produced
superimposing $J/\psi$ and $Z$ events from PYTHIA~\cite{Sjostrand:2006za} into
simulated \PbPb events generated with the HIJING~\cite{Wang:1991hta}
event generator.  HIJING was run in a mode with effects from jet quenching turned off,
since they have not been adjusted to agree with existing experimental data.  Elliptic flow
was imposed on the events subsequent to generation, with a magnitude and
$p_{\mathrm{T}}$ dependence derived from RHIC data.
The detector response to the complete PYTHIA+HIJING event is 
simulated~\cite{ATLASSim} using GEANT4~\cite{Agostinelli:2002hh}. 

Lead-lead collision centrality percentiles
are defined from the total 
transverse energy, $\Sigma E_{\mathrm{T}}^{\mathrm{FCal}}$,
measured in the forward calorimeter (FCal), which
covers $3.2 < |\eta| <4.9$. 
The same conventions and bins for centrality are used as in our previous publication~\cite{ATLASjetq}.
The centrality
dependence of the muon detection efficiency is parameterized as
a function of the total number of hits per unit of pseudorapidity 
detected in the first pixel layer. This is strongly correlated
to $\Sigma E_{\mathrm{T}}^{\mathrm{FCal}}$, but gives a more
direct measure of the ID occupancy.
The full data sample is divided into four bins of collision
centrality, 40-80\%, 20-40\%, 10-20\%, and 0-10\%. The most peripheral
20\% of collisions are excluded from this analysis due to larger
systematic uncertainties in 
estimating the number of binary 
nucleon-nucleon collisions in these events.

The $J/\psi \to \mu^{+} \mu^{-}$ reconstruction efficiency is obtained from the
MC samples as a function of the event centrality. 
The inefficiency gradually increases from peripheral to central 
collisions, due primarily to
an occupancy-induced inefficiency in the ID tracking, as shown in 
Table~\ref{TableYield}.
The $Z \to \mu^{+} \mu^{-}$  
reconstruction efficiency is obtained in a similar way.

\begin{figure*}[t]
\begin{center}
\includegraphics[width=16pc]{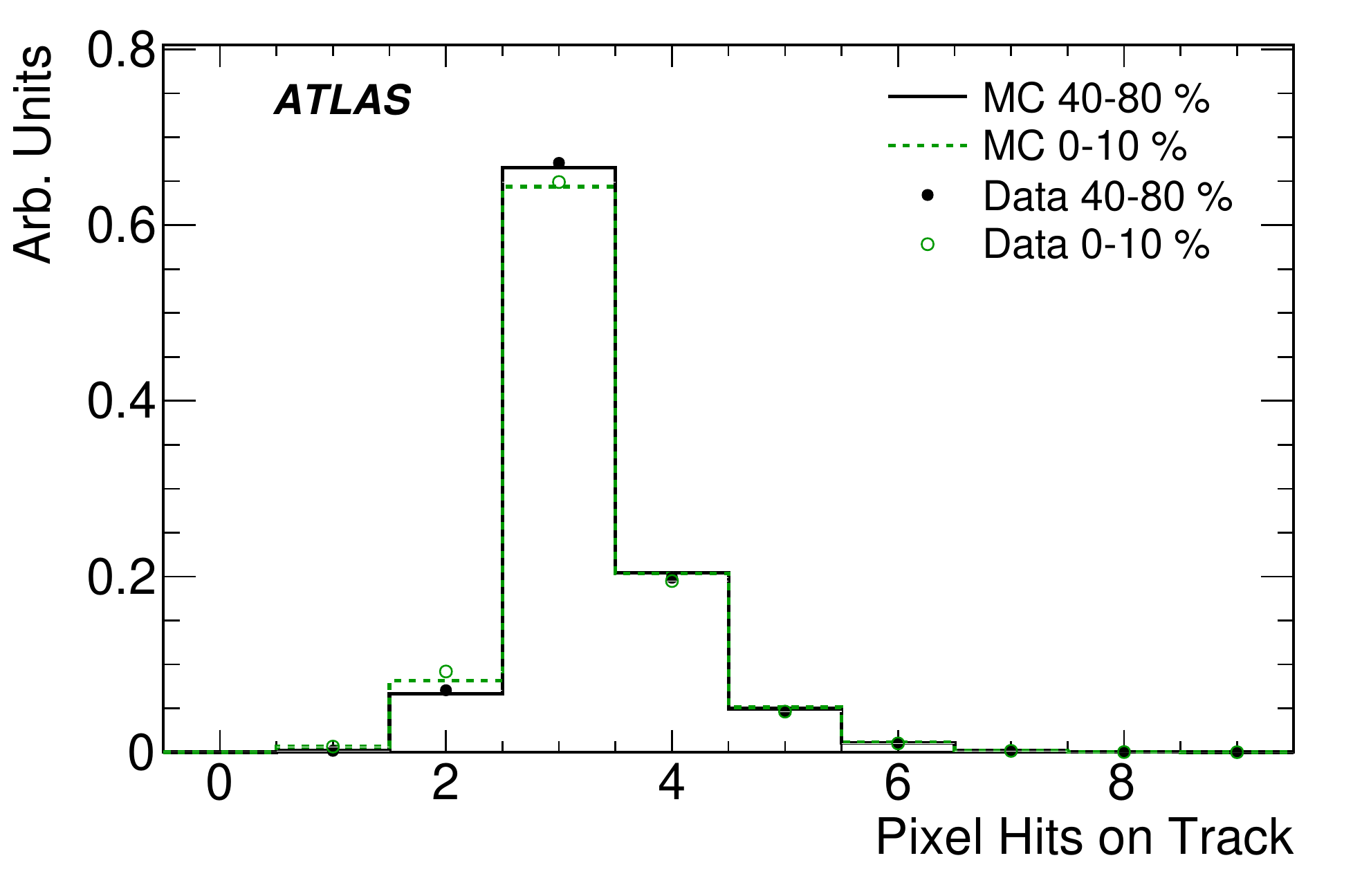}
\includegraphics[width=16pc]{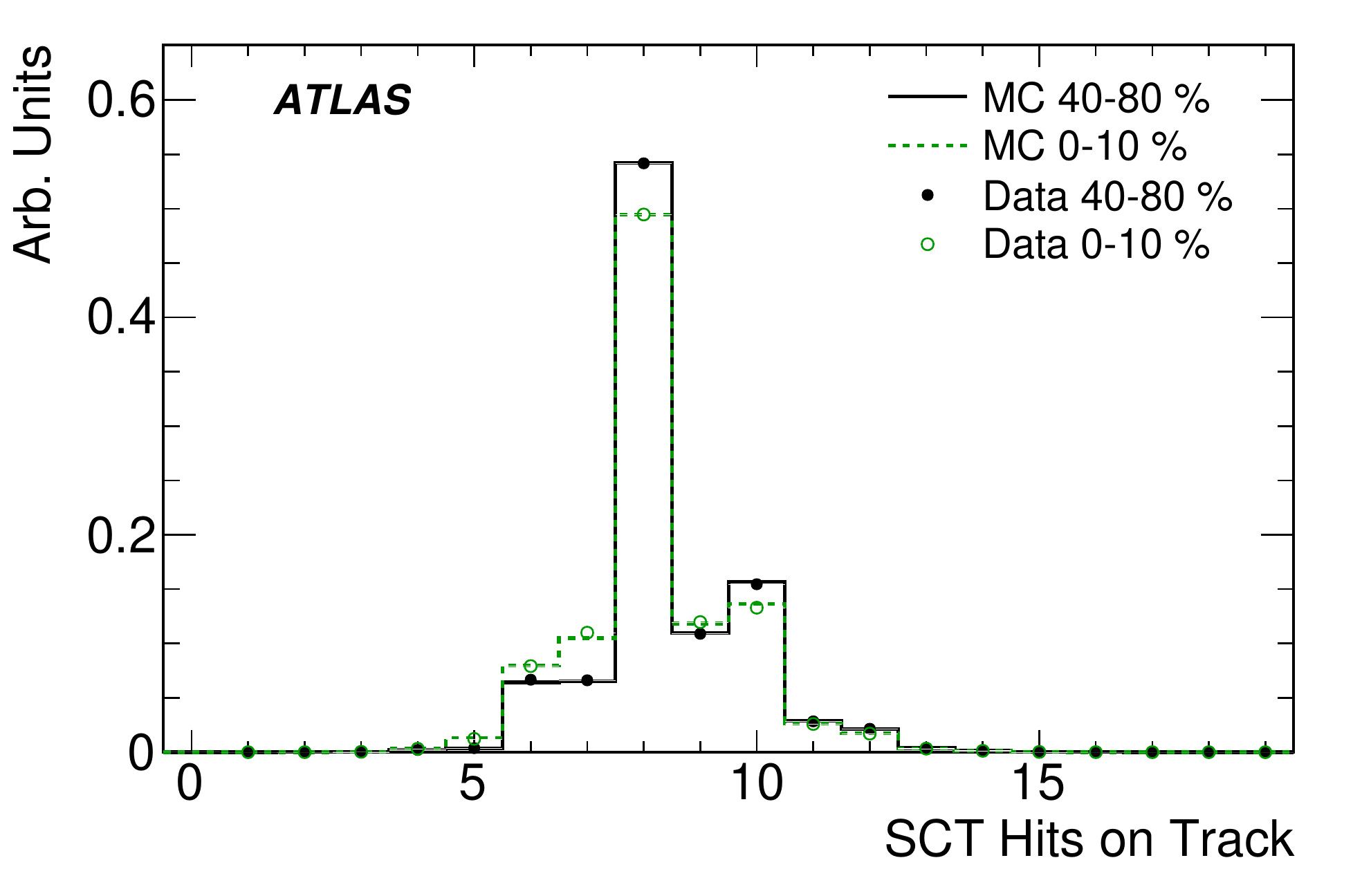}
\includegraphics[width=16pc]{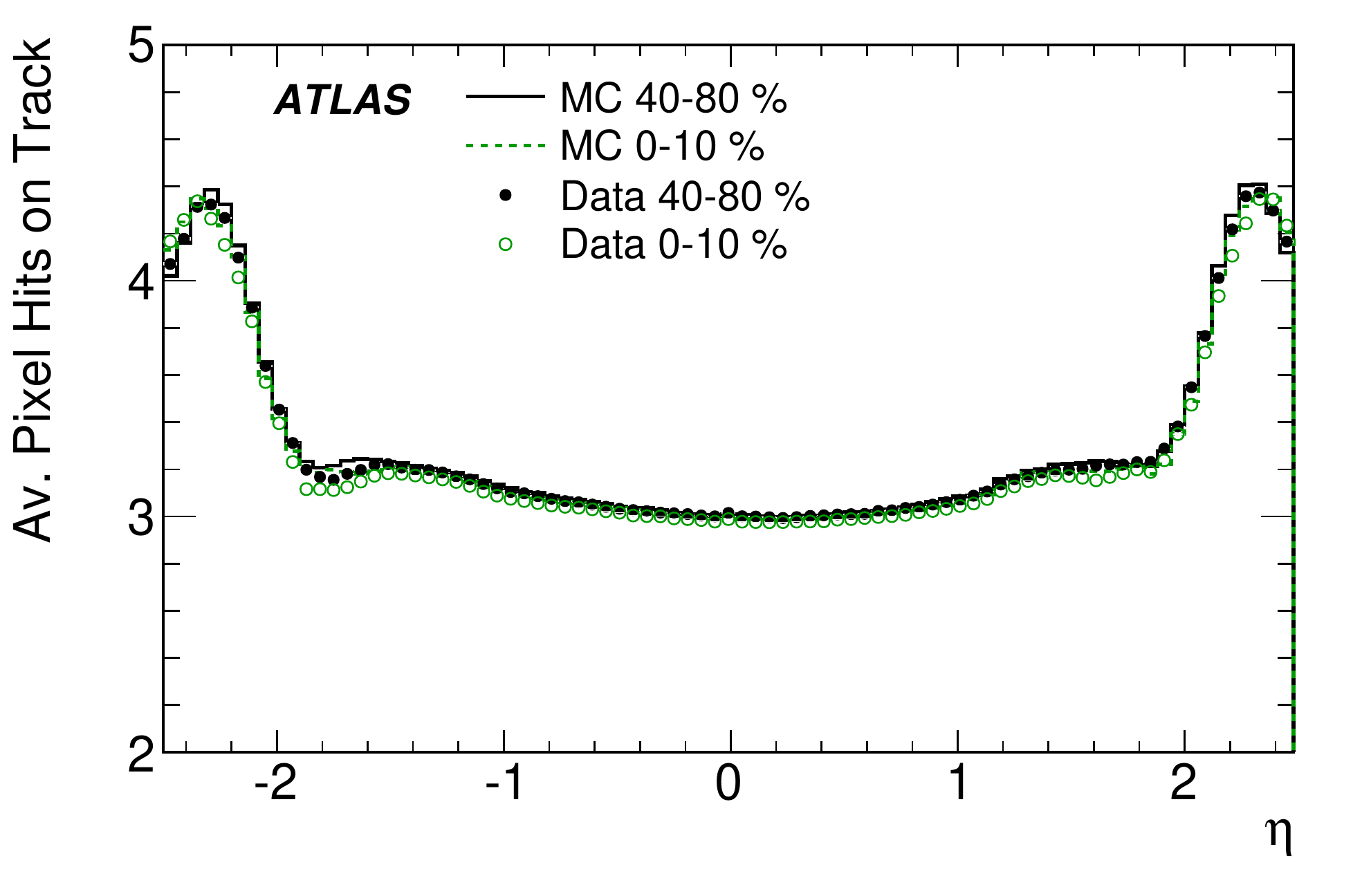}
\includegraphics[width=16pc]{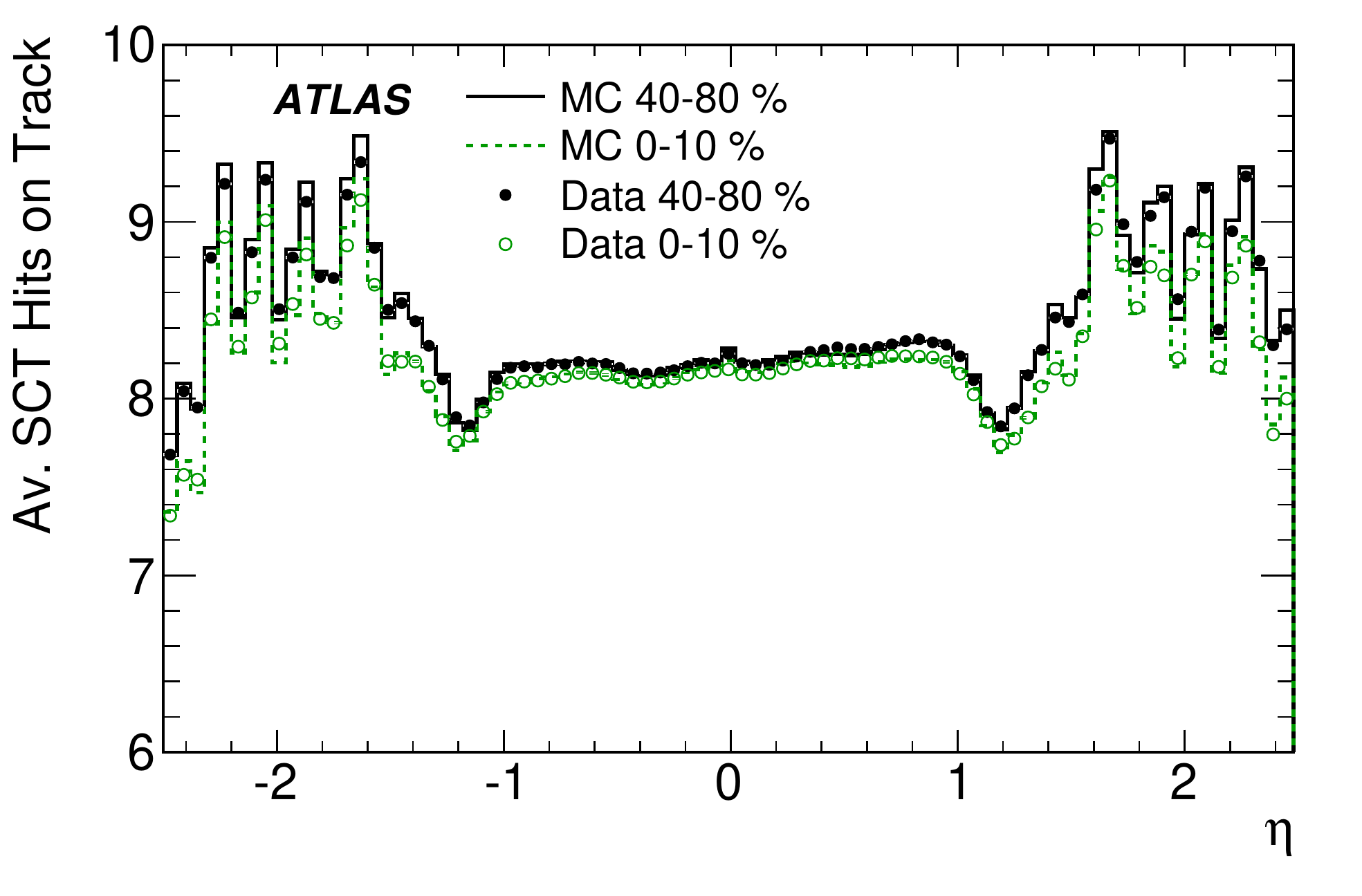}
\caption{\label{hitsID}
(top row) The number of Pixel (left) and SCT (right) hits on tracks for data (points with errors) and MC (histogram) for two different centrality bins: 0-10\% (open/dotted) and 40-80\% (closed/solid).
(bottom row) The average number of Pixel (left) and SCT (right) hits as a function of $\eta$ for MC and data in the same two centrality bins.
}
\end{center}
\end{figure*}

An example of the very good agreement between data and MC in different centrality bins is presented 
in Figure~\ref{hitsID},
which shows the numbers of Pixel and SCT hits associated to tracks selected with a 
looser $p_{\mathrm{T}} >0.5$ GeV cut than that for the $J/\psi$. 
The figure shows results for data and MC at two different centralities (0--10\% and
40--80\%). The distributions of the number of hits averaged over $\eta$
and the average number of hits as a function of $\eta$ are shown.
The slight decrease of the number of SCT hits on track as a function of centrality 
is well reproduced by the simulation, demonstrating that the dense 
environment of the most central collisions is reasonably well modelled.

\section{$J/\psi$ production as a function of centrality}
\label{Signals}

\begin{figure*}[t]
\begin{center}
\includegraphics[width=16pc]{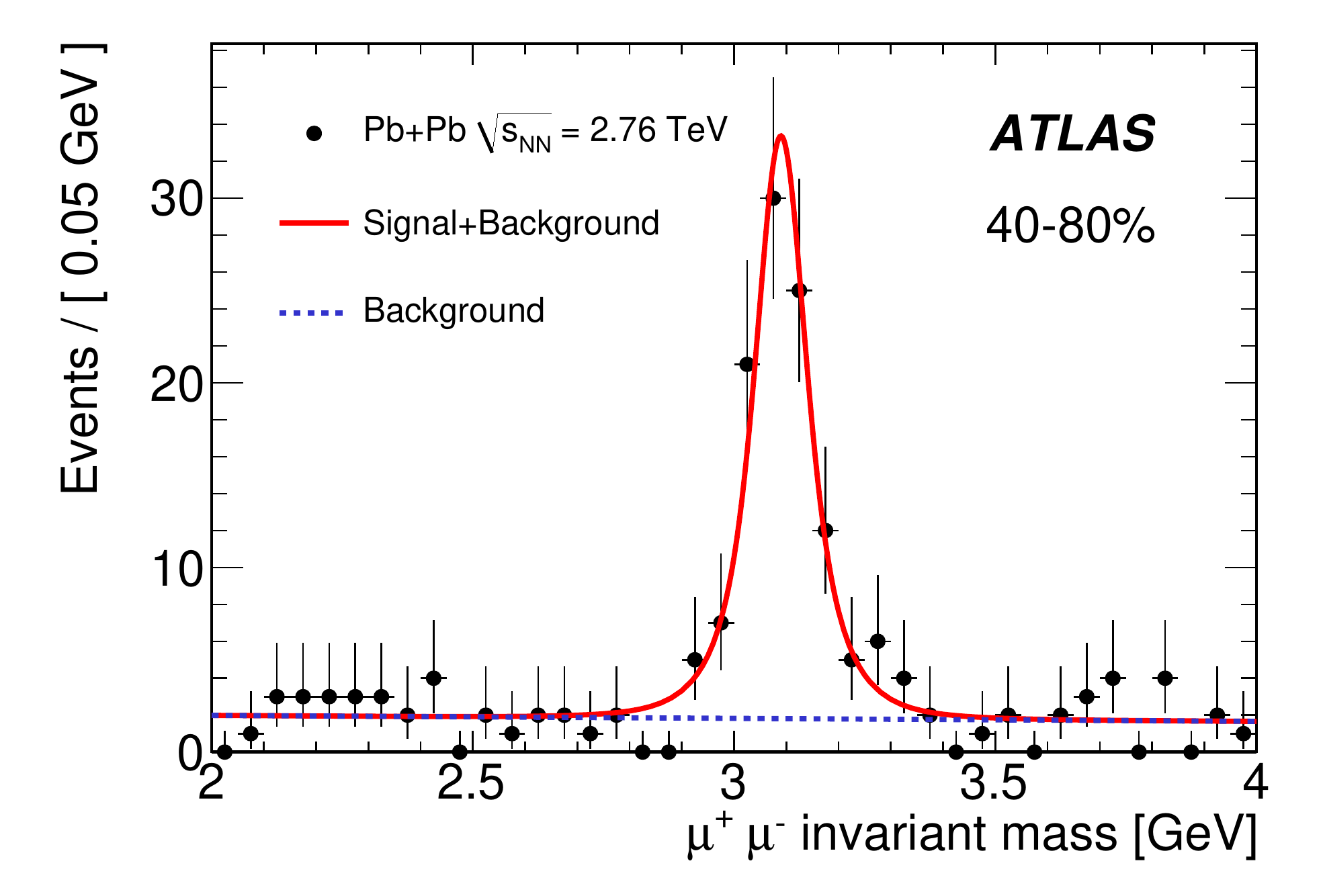}
\includegraphics[width=16pc]{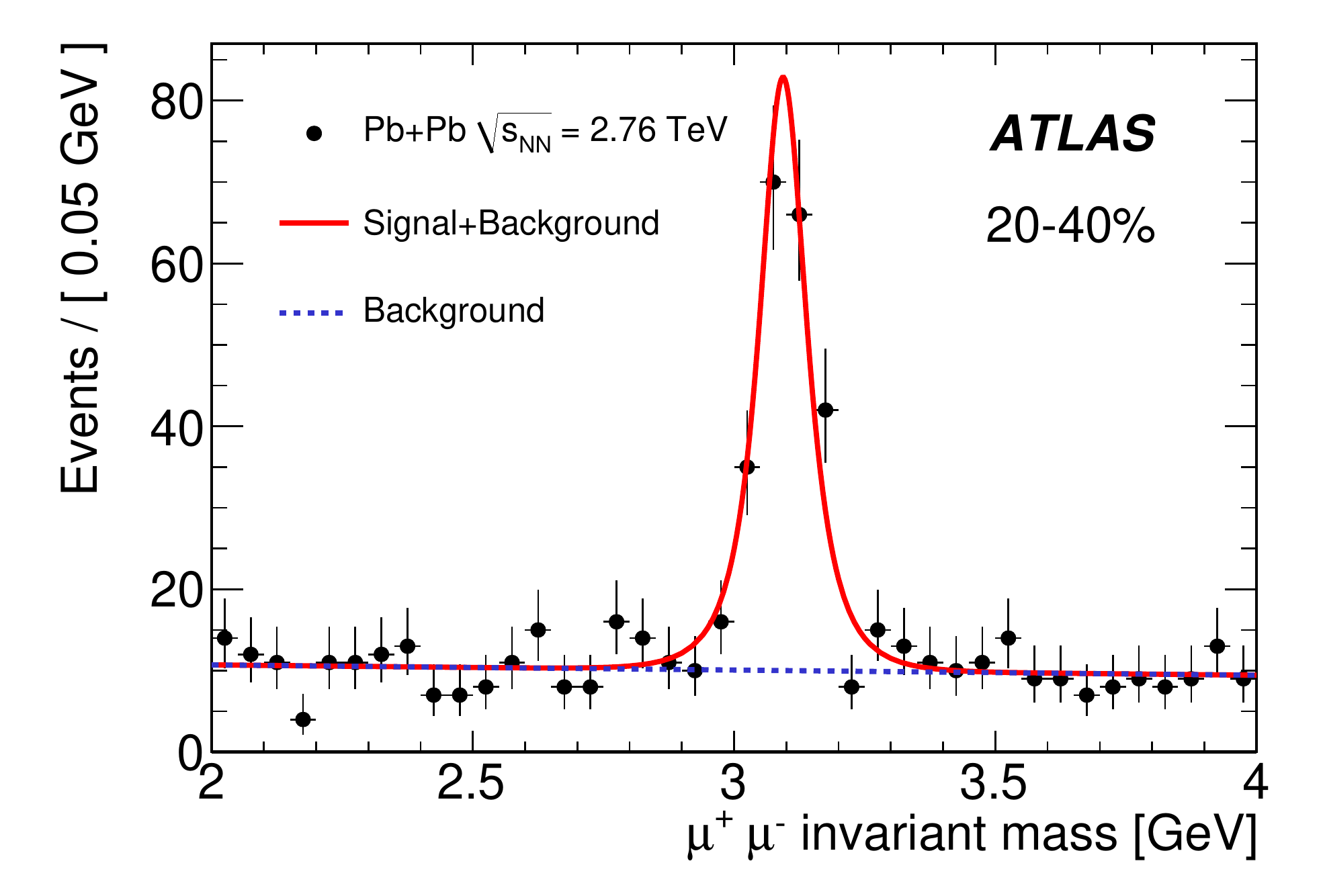}
\includegraphics[width=16pc]{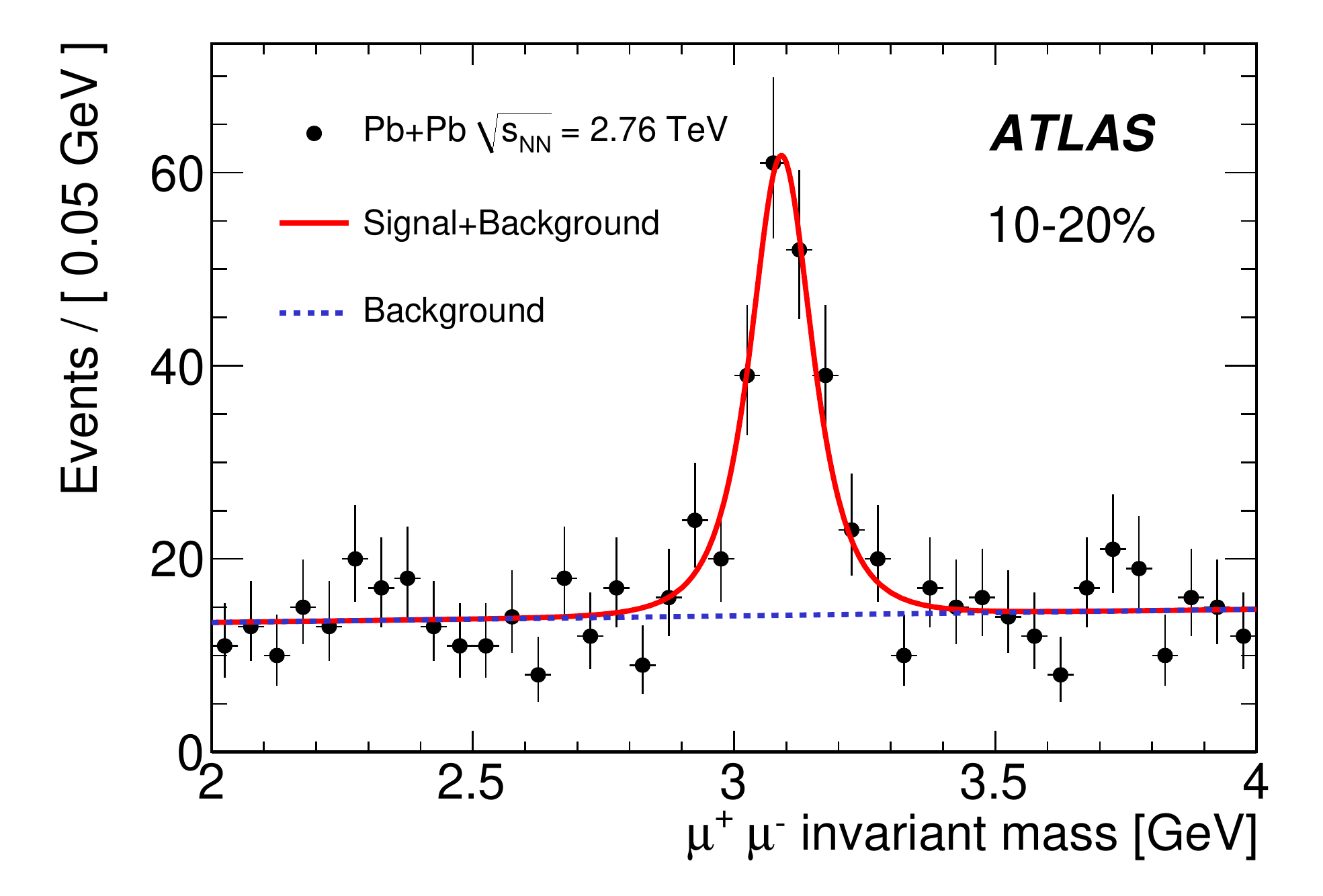}
\includegraphics[width=16pc]{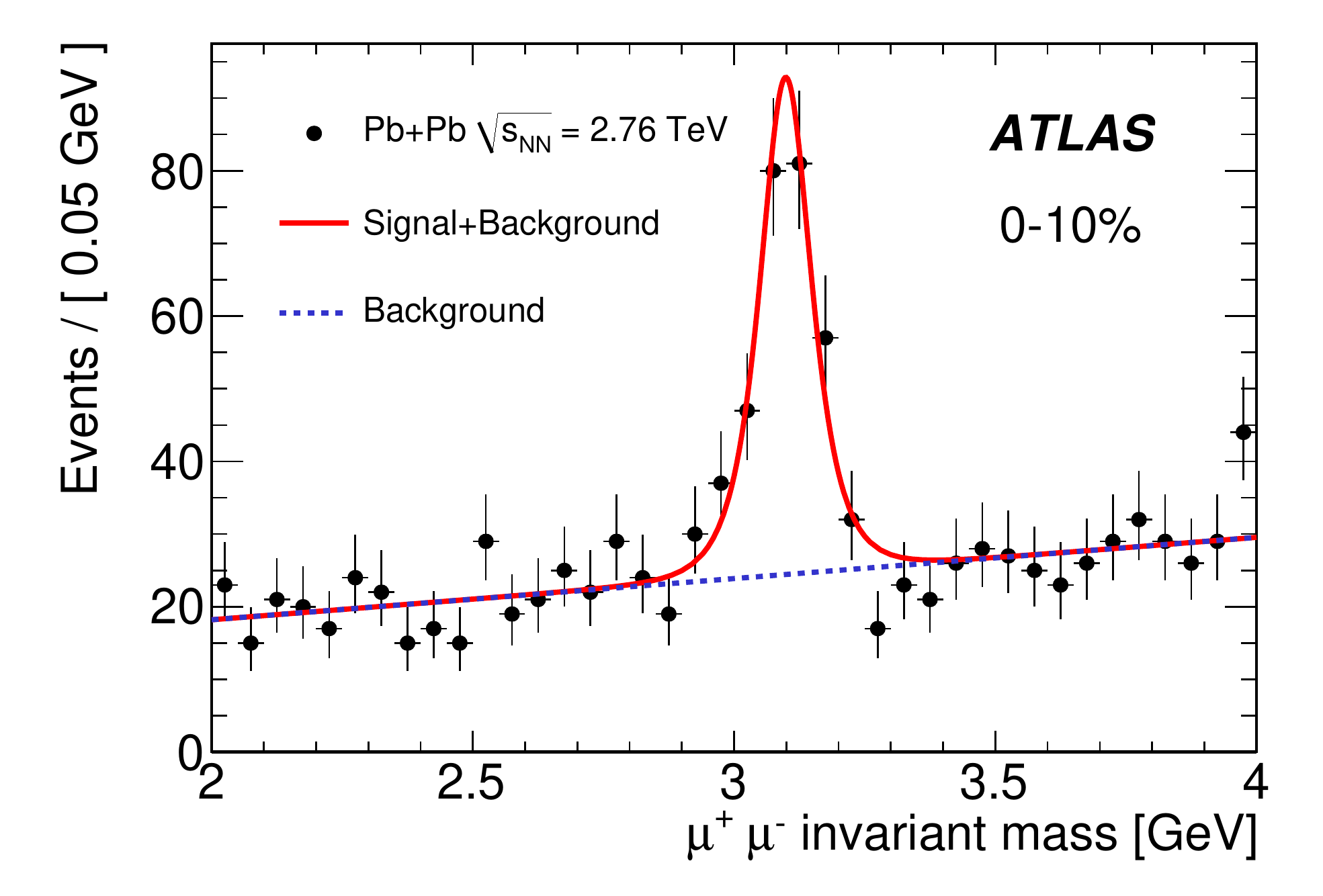}
\caption{
\label{mass}
Oppositely-charged di--muon invariant mass spectra in the four considered
centrality bins from most peripheral (40-80\%) to most central (0-10\%).
The $J/\psi$ yields in each centrality bin are 
obtained using a sideband technique. The fits 
shown here are used as a cross check.
}
\end{center}
\end{figure*}

\begin{table}[bth]
\centering
\begin{tabular}{| c | c | c | c | c | c |}
\hline
Centrality & $N^\mathrm{meas}(J/\psi)$ & $ \epsilon (J/\psi)_c / $ & \multicolumn{3}{|c|}{Systematic Uncertainty}\\
         &  &        $ \epsilon (J/\psi)_{40-80}$   & Reco. eff.    & Sig. extr. &  Total  \\
\hline
0-10\%  & 190 $\pm$ 20 & 0.93 $\pm$ 0.01 & 6.8 \% & 5.2 \% & 8.6 \%  \\
10-20\% & 152 $\pm$ 16 & 0.91 $\pm$ 0.02 & 5.3 \% & 6.5 \% & 8.4 \%  \\
20-40\% & 180 $\pm$ 16 & 0.97 $\pm$ 0.01 & 3.3 \% & 6.8 \% & 7.5 \%  \\
40-80\% & \phantom{0}91 $\pm$ 10 &   1   & 2.3 \% & 5.6 \% & 6.1 \%  \\
\hline
\end{tabular}
\caption{
The measured numbers of $J/\psi$ signal events per centrality bin before any correction,
with their statistical
errors, are listed in the second column.  The relative efficiency
corrections derived from the simulation are also shown, with the 
MC statistical error. The last columns give
the experimental systematic uncertainties 
on the reconstruction efficiency and signal extraction, as well as the
total uncertainty.}
\label{TableYield}
\end{table}  

The oppositely-charged di--muon invariant mass spectra 
in the $J/\psi$ region after the selection are shown in Figure~\ref{mass}.
The number of $J/\psi \to \mu^{+} \mu^{-}$ decays is then found by a
simple counting technique. The signal mass window is defined by the range 2.95--3.25~GeV.
The background is derived from two mass sidebands, 2.4--2.8~GeV and 3.4--3.8~GeV, 
with a linear extrapolation.
To determine the uncertainties related to the signal extraction,
an alternative method based on a maximum likelihood fit with the 
mass resolution left as a free parameter is used as a cross check,
as explained in section~\ref{ExpSyst}. 
Centrality-dependent efficiency corrections, derived from Monte Carlo events, are
applied to the resulting signal yields.
The number of $J/\psi$ decays after background
subtraction, but before any other correction, 
are listed in Table~\ref{TableYield}.
With the chosen transverse momentum cuts on the decay muons, 80\% of the reconstructed
$J/\psi$ have $p_T>6.5$ GeV.

The measured $J/\psi$ yields at different centralities are corrected by
the reconstruction efficiency $\epsilon_c$ for $J/  \psi \to \mu^{+} \mu^{-}$,
derived from MC and parameterized in each centrality bin,
and 
the width of the centrality bin, $W_c$, which represents a well-defined fraction of the minimum bias events.
The corrected yield of $J/\psi$ mesons is given by:
\begin{equation}
N_c^{\mathrm{corr}} (J/  \psi \to \mu^{+} \mu^{-})= \frac{ N^{\mathrm{meas}}(J/  \psi \to \mu^{+} \mu^{-} )_c }
{\epsilon (J/\psi)_c \cdot W_c}.
\label{eq:ratiocent}
\end{equation}
The ``relative yield'' is defined by normalizing to the yield found in the 
most peripheral 40-80\% 
centrality bin: $R_c = N_{c}^{\mathrm{corr}}/{N^{\mathrm{corr}}_{40-80\%}}$.
Note that the uncertainties in the 40-80\% bin are not propagated into this ratio for the more central bins.
Finally, the ``normalized yield'' is defined by scaling the relative yield by
the ratio $R_{\mathrm{coll}}$ of the mean number of 
binary collisions $N_{\mathrm{coll},c}$, detailed in section \ref{Rcoll}, in each centrality bin to that for the most
peripheral (40-80\%) bin: $R_{cp} = R_{c}/R_{\mathrm{coll}}$.

\subsection{Experimental systematic uncertainties}
\label{ExpSyst}

Several experimental 
systematic effects are considered. These are grouped into
those affecting the $J/\psi$ reconstruction efficiency,
and those from the extraction of the number of signal events
from the di--muon mass spectra.  Since this measurement only 
determines the relative yields as a function of centrality, only the centrality dependence of these 
effects is relevant. Any uncertainty on the absolute value 
cancels out in the ratio. The variation
of the $J/\psi$ reconstruction efficiency with centrality observed in simulation 
is mainly due to the larger occupancy in the ID. 
Because of the low occupancy in the MS by the primarily-soft tracks produced
in heavy ion collisions, the fraction of
muons from $J/\psi$ decays with a reconstructed track in the MS is independent of
centrality within the MC statistical uncertainty.
On the other hand, to improve the reliability of the ID track reconstruction in the 
dense environment, 
rather stringent track quality requirements are made, relative to those
defined for proton-proton collisions~\cite{ATLASminb}. 
In particular, there must be at least nine silicon
hits on each track, with no missing pixel hits and 
not more than one missing SCT hit, 
in both cases where such hits are expected.
In order to evaluate systematic uncertainties,
comparisons have been made between the distributions of hits 
associated with tracks and missing
hits between data and MC as a function of centrality.
The differences between the fraction of tracks with associated or missing hits
close to the track selection cuts have been used to derive the
systematic uncertainties on the ID track reconstruction that range between
1 and 3\% as a function of the centrality.
These uncertainties are fully correlated for both muons 
from the $J/\psi$ decay, resulting in a systematic uncertainty up to
about 7\% on the $J/\psi$ reconstruction efficiency.
As an additional cross-check, the ID reconstruction
was run with looser cuts on the number of missing pixel and SCT hits,
in order to study directly the number of tracks lost because of the cuts 
on these quantities.
The resulting track losses, as a function of centrality in data and simulation, 
were compatible with the systematic uncertainties derived with the hit 
comparison method described above.
Further cross-checks have been made by studying 
the matching between the MS and ID momentum measurements, and
by examining variables
such as the track multiplicity distribution in a cone of $\Delta R < 0.1$ 
(where $\Delta R^2 = \Delta \phi^2 + \Delta \eta^2$) around muon candidates, 
and by evaluating the relative momentum difference between the two independent 
measurements of the same muon candidate. 
The fraction of muons measured in the MS but not matched to any ID 
track has also been compared in data and MC as a function of centrality.
All of these studies show that
the MC reproduces well the behaviour of the data as a function of centrality. 
The relative
statistical uncertainty on the MC efficiency corrections ranges between 1.6 and 3.2\%
and this is combined in quadrature with the other uncertainties.

To address the uncertainties associated with the $J/\psi$ signal extraction, an
independent method based on an unbinned maximum likelihood fit is
used to evaluate the number of signal events from the di--muon mass
spectra. An overall scale factor on the event-by-event
mass resolution is a free parameter of the fit, allowing for 
possible variations of resolution with centrality.
Two different background parameterizations are used, with either a first or second order polynomial.
The maximum deviation of the fitted yield compared to the 
sideband subtraction method is taken as the systematic
uncertainty on the signal extraction.

The systematic uncertainties from the different sources are listed in 
Table~\ref{TableYield}.

\subsection{Definition of $N_{\mathrm{coll}}$}
\label{Rcoll}

The mean number of binary nucleon-nucleon collisions, $N_{\mathrm{coll}}$,
corresponding to each centrality bin was calculated using a
Glauber Monte Carlo package that has been applied
extensively at RHIC energies~\cite{Alver:2008aq, Miller:2007ri}. 
The impact parameter is selected randomly event by event, and both the number of
participating nucleons which undergo at least one inelastic collision
($N_{\mathrm{part}}$) and the number of binary collisions (i.e.
the total number of nucleon-nucleon collisions,
$N_{\mathrm{coll}}$) are calculated for each event.
The primary experimental inputs to the
Glauber calculation are the radius 
($R$) and skin depth ($a$) parameters of the Wood-Saxon
parameterization of the nuclear density 
($\rho(r) = \rho_0/[1+\exp((r-R)/a)]$), $R = 6.62
\pm 0.06~\mathrm{fm}$ and $a = 0.546 \pm  0.010~\mathrm{fm}$, 
respectively~\cite{De Jager:1987qc}, 
and the nucleon-nucleon
inelastic cross-section, assumed to be $\sigma_{\mathrm{inel}} = 64 \pm 6$~mb
from an extrapolation of lower energy data.
Using these parameters, the Glauber calculations give a total inelastic cross section of
7.6 barns, which is defined as the ``geometric'' cross section below.

Systematic uncertainties on the resulting $R_{\mathrm{coll}}$ values are 
estimated by separately varying $R$, $a$ and $\sigma_{\mathrm{inel}}$ 
by one standard deviation. 
The variations of $R$ and $a$ are found to give
results of the same magnitude but opposite sign, 
indicating that the uncertainties on the two
parameters are correlated.  However, 
they are conservatively treated 
as uncorrelated for the error analysis used in these studies.

Any possible variation in the
fraction of the geometric cross section selected by
the combination of trigger and event selection criteria,
$\varepsilon_{\mathrm{mb}}$, 
as a function of centrality 
must also be considered in evaluating
systematic uncertainties on the lead-lead collision geometry,
so that the centrality percentiles correspond to the 
correct fractions of the
efficiency-corrected geometric cross section.
The uncertainty is estimated by examining
the distribution of $\Sigma E_{\mathrm{T}}^{\mathrm{FCal}}$ 
in the independent data sample 
selected by a random trigger and
filtered by requiring a minimal amount of Inner Detector activity.
The event selection
criteria described above are also applied,
with an additional requirement that both ZDCs see energies
consistent with the presence of at least one neutron. 
This combination of vertex, MBTS and ZDC 
selections efficiently rejects photonuclear
interactions~\cite{Djuvsland:2010qs}.
The total selected fraction of the geometric cross section is estimated using a
fit to the resulting $\Sigma E_{\mathrm{T}}^{\mathrm{FCal}}$ 
distribution, assuming the transverse energy in
each event results from a superposition of participating nucleons and 
binary collisions (a similar assumption to that used in Ref.~\cite{Collaboration:2010cz}):
\begin{equation}
\Sigma E_{\mathrm{T}}^{\mathrm{FCal}} = E_{\mathrm{T}}^{pp} \left\{ (1-x)\frac{N_{\mathrm{part}}}2 + xN_{\mathrm{coll}} \right\}.
\end{equation}
In this formula, $E_{\mathrm{T}}^{pp}$ is
the value of $\Sigma E_{\mathrm{T}}^{\mathrm{FCal}}$ 
when $N_{\mathrm{part}}=2$ and $N_{\mathrm{coll}}=1$ 
(the values for a single proton-proton collision) and 
$x$ controls the relative contribution of 
participants and binary collisions in lead-lead events. 
An additional 
constant noise term is also included
to account for the low energy part of the
distribution.
Distributions of $\Sigma E_{\mathrm{T}}^{\mathrm{FCal}}$ are
generated for 500k MC events and fitted to the data for 
a range of values of $x$ (from 0.09 to 0.15), and also varying
$E_{\mathrm{T}}^{pp}$ and the noise term.
For all cases, the integral of the observed
distribution in data accounts for around $98\%$ of the best fit to the 
simulated distribution,
with a variation of around $1\%$. 
This provides an estimate of the total event selection efficiency
$\varepsilon_{\mathrm{mb}}$ 
relative to the geometric cross section.
An absolute systematic error of $\pm 2\%$ is assigned to
$\varepsilon_{\mathrm{mb}}$ with the positive range also accounting for
the possible leakage of photonuclear events into the event sample used to
obtain the $\Sigma E_{T}^{\mathrm{FCal}}$ distribution.

\begin{table}[th]
\centering
\begin{tabular}{| c | c | c | }
\hline
Centrality & $R_{\mathrm{coll}} $       & Uncertainty \\
\hline
0-10\%   & 19.5  &  5.3 \% \\
10-20\%  & 11.9  &  4.7  \%  \\
20-40\%  &  5.7  &  3.2 \% \\
40-80\%  &  1.0  &  -- \\
\hline
\end{tabular}
\caption{
The correction factors
$R_{\mathrm{coll}}$, together with the relative systematic uncertainty, 
stated as a 1$\sigma$ value.}
\label{TableRcoll}
\end{table}  

The total systematic uncertainties on the ratios
$R_{\mathrm{coll}}$ are evaluated by combining the variations with $R$, $a$,
$\sigma_{\mathrm{inel}}$ and $\varepsilon_{\mathrm{mb}}$, in quadrature.
The values of $R_{\mathrm{coll}}$ and their systematic uncertainties are
reported in Table~\ref{TableRcoll}.
It should be noted that the estimate of $\varepsilon_{\mathrm{mb}}$ 
leads to correlations between the extracted
values of $N_{\mathrm{coll}}$, and thus the uncertainties on $R_{\mathrm{coll}}$ are
also correlated bin-to-bin.

\subsection{$J/\psi$ yields}

\begin{figure*}[t]
\begin{center}
\includegraphics[width=16pc]{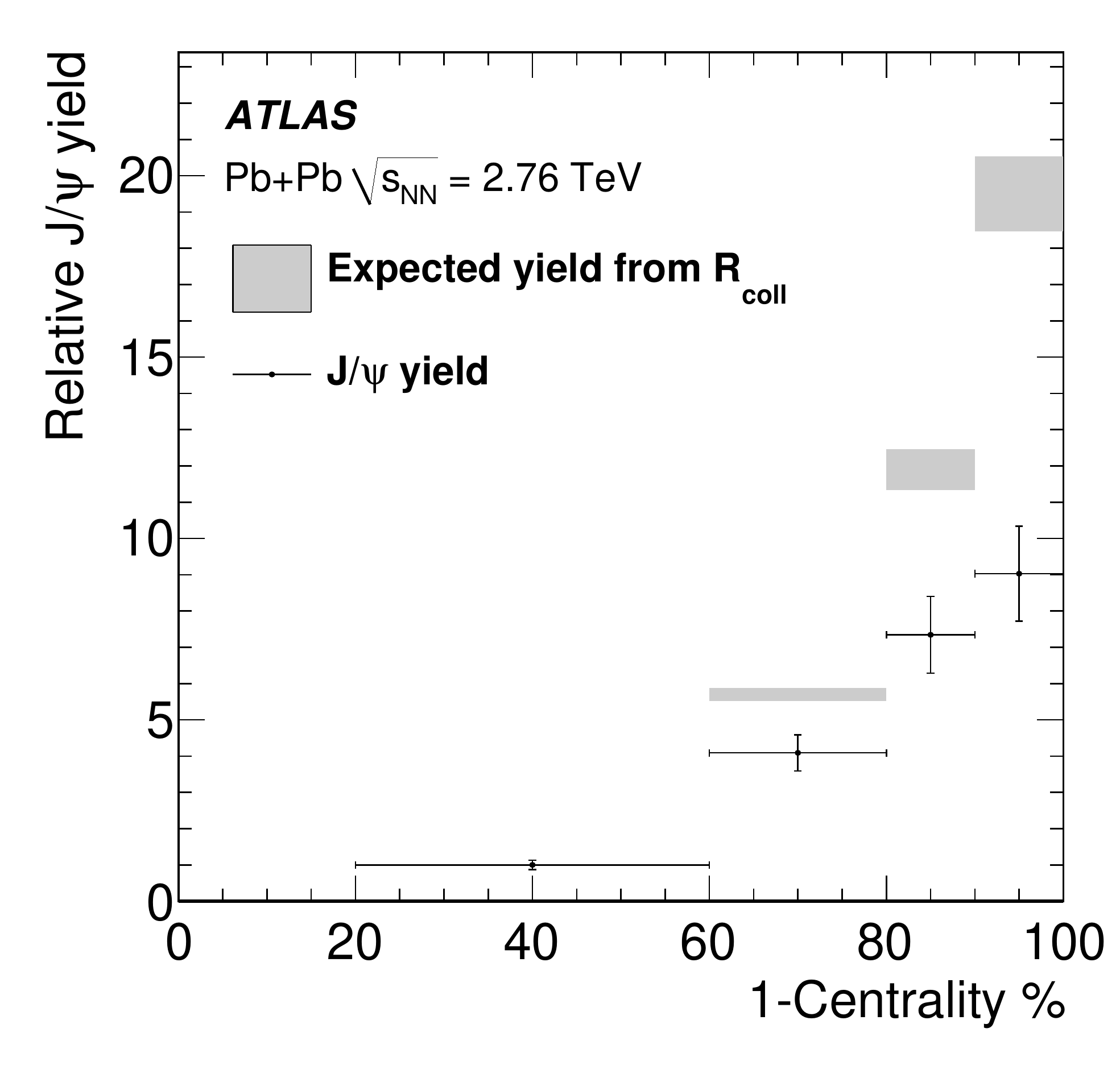}
\includegraphics[width=16pc]{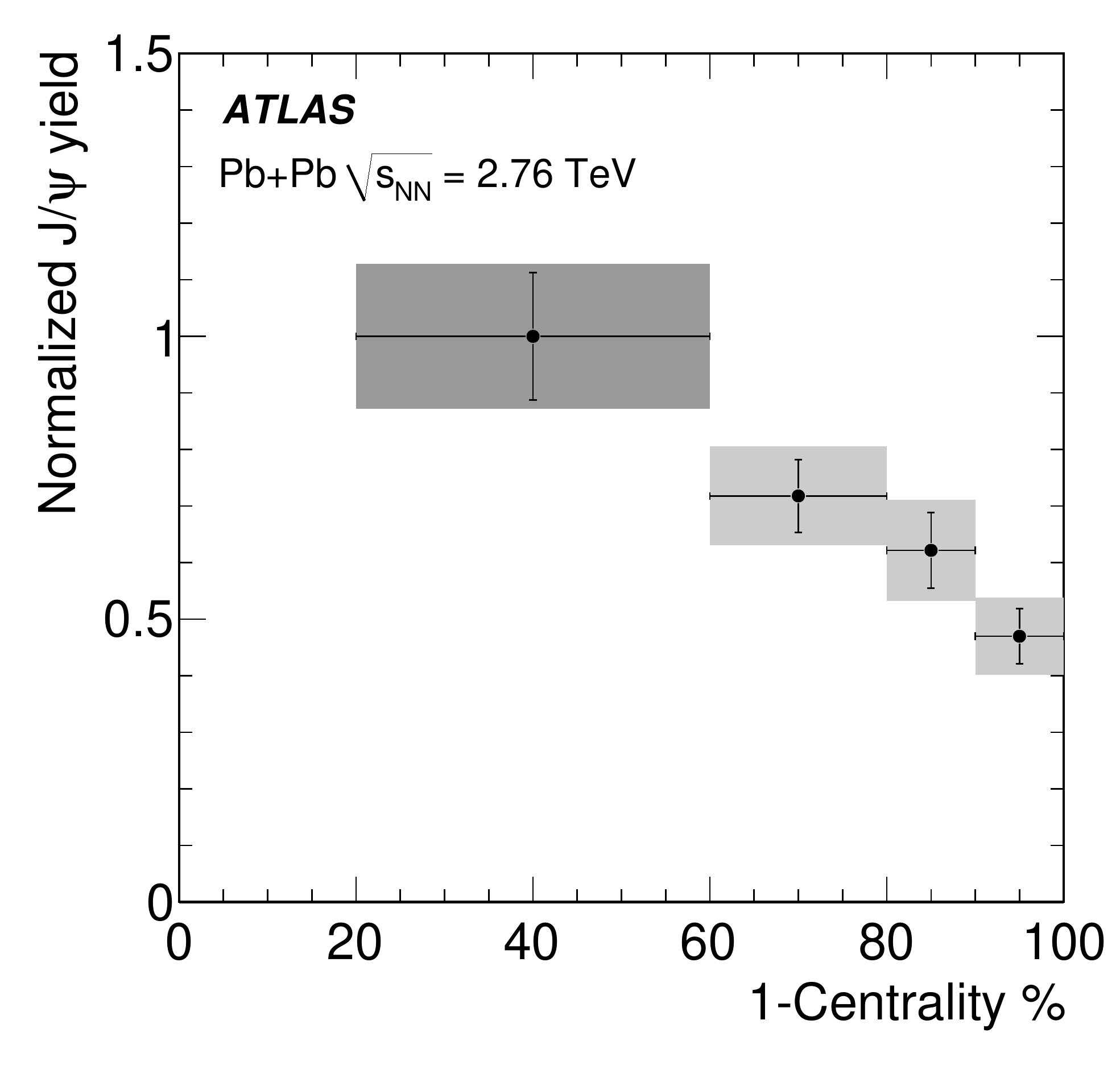}
\caption{ \label{yields}
(left) Relative $J/\psi$ yield as a function of centrality normalized to the most peripheral bin (black dots with errors).  
The expected relative yields
from the (normalized) number of binary collisions ($R_{\mathrm{coll}}$) are also
shown (boxes, reflecting 1$\sigma$ systematic uncertainties).
(right) Value of $R_{cp}$, as described in the text, as a function of centrality. 
The statistical errors are shown as vertical 
bars while the grey boxes also include the combined systematic errors.
The darker box indicates that the 40-80\% bin is used to set the scale for all bins, but the uncertainties in this bin are not propagated into the more central ones.
 }
\end{center}
\end{figure*}

The relative $J/\psi$ yields after normalization and efficiency corrections 
as in equation ~\ref{eq:ratiocent}, $R_c$, are 
compared to the expected $R_{\mathrm{coll}}$ values in the left panel of Figure~\ref{yields}. 
The yield errors are computed by adding the statistical and systematic uncertainties in quadrature.
A clear difference is observed as a function of centrality between
the measured relative $J/\psi$ yield and the prediction based on
$R_{\mathrm{coll}}$, 
indicating a deviation from the simplest expectation based on QCD factorization.
The ratio of these two values, $R_{cp}$, is shown as a function of centrality in
the right panel of Figure~\ref{yields}.
The data points are not consistent with their average, 
giving a $P(\chi^{2},N_{\mathrm{DOF}})$ value of 0.11\% with three
degrees of freedom, computed conservatively ignoring any correlations among 
the systematic uncertainties.  Instead, a significant decrease of $R_{cp}$ as a 
function of centrality is observed.

\section{$Z$ production as a function of centrality}
\label{Zprod}

\begin{table}[t]
\centering
\begin{tabular}{| c | c | c |}
\hline
Centrality & N($Z$) & $ \epsilon (Z)_c / \epsilon (Z)_{40-80}$   \\
\hline
0-10\%  & 19             & 0.99 $\pm$ 0.01 \\
10-20\% & \phantom{0}5   & 0.97 $\pm$ 0.01 \\
20-40\% & 10             & 0.98 $\pm$ 0.01 \\
40-80\% & \phantom{0}4   &   1             \\
\hline
\end{tabular}
\caption{
The number of $Z$ events per centrality bin 
and the relative efficiency
corrections derived from the simulation.}
\label{TableZ}
\end{table}  

\begin{figure*}[t]
\begin{center}
\includegraphics[width=16pc]{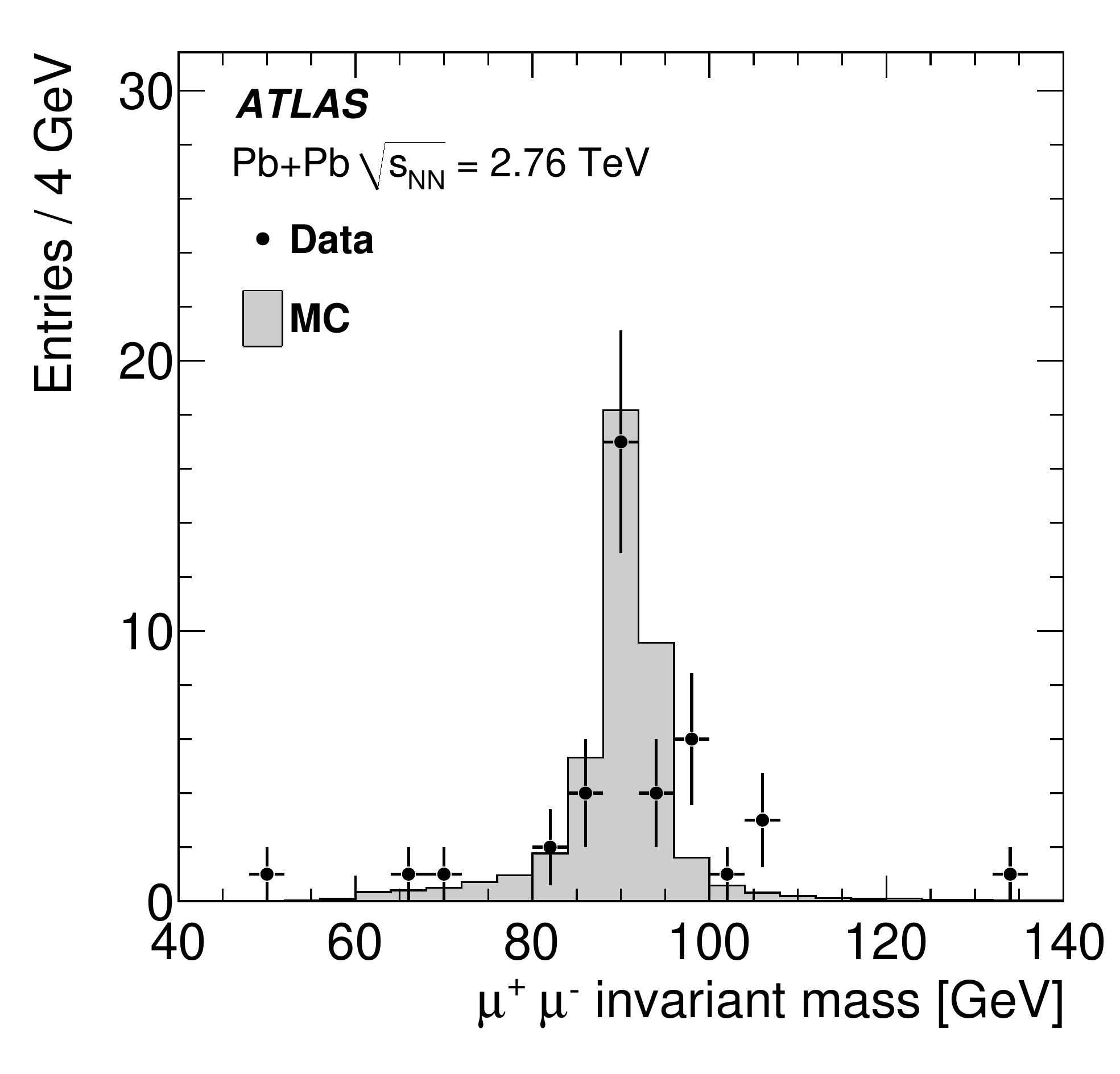}
\includegraphics[width=16pc]{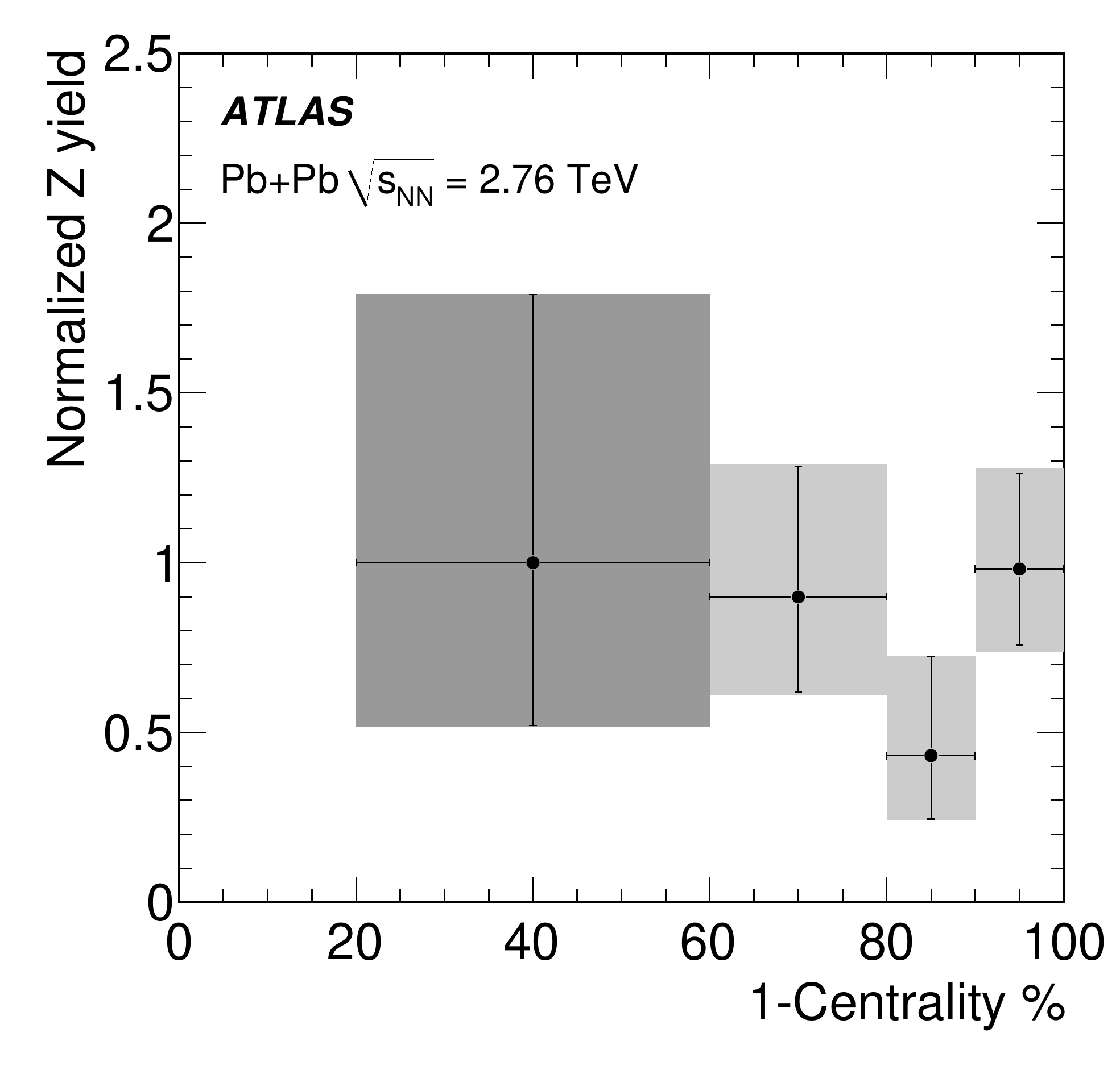}
\caption{
\label{z}
The di--muon invariant mass (left) after the selection described in the text.
The value of $R_{cp}$ (right) computed with the 38 selected $Z$ candidates.
The statistical errors are shown as vertical 
bars while the grey boxes also include the combined systematic errors.
The darker box indicates that the 40-80\% bin is used to set the scale for all bins, but the uncertainties in this bin are not propagated into the more central ones.
}
\end{center}
\end{figure*}

$Z$ candidates are selected by requiring a pair of oppositely charged
muons with $p_{\mathrm{T}} > 20$~GeV and $|\eta| < 2.5$~\cite{zcspaper}.
An additional cosmic ray rejection cut on the sum of the
pseudorapidities of the two muons,
$|\eta_1+\eta_2| > 0.01$, is 
also applied.
The invariant mass distribution of the selected pairs is shown in 
the left panel of Figure~\ref{z}. 
With this selection, 38 $Z$ candidates are retained in the signal mass window of 66 to 116 GeV.
The background after this selection is expected to be below 2\%, 
and is not corrected for in the result. 
The number of $Z$ events
in each centrality bin is given in Table~\ref{TableZ}.

The $R_{cp}$ variable for the $Z$ 
candidates is computed in the same way as for the $J/\psi$ sample.
The relative efficiency corrections determined from dedicated MC samples are
given in Table~\ref{TableZ}.
For high transverse momentum tracks, the reconstruction is expected to perform as well 
as or better than in the low $p_{\mathrm{T}}$ regime characteristic of the $J/\psi$ study.
For this reason, the same systematic uncertainties as for the $J/\psi$ results have 
been applied to the $Z$ relative yield measurements.  Several cross-checks have been performed to 
support this assumption. In addition to the tracks reconstructed with the combined 
ID and MS information,
tracks reconstructed by the MS alone have been checked, 
and only one additional candidate was found.
This candidate has been inspected and an ID track was in fact 
found but with too few hits to pass the stringent reconstruction requirements. 
The $Z$ selection was also applied to same charge muon pairs, and no 
candidates were selected within the 66--116~GeV mass window.
To control the residual background from cosmic rays, the distribution of the difference of 
the transverse impact parameters of the two muons from $Z$ candidates 
was examined and found to be compatible with that expected for collision muons.

The measured $Z$ yields are displayed in the right panel of 
Figure~\ref{z}, normalized to the yield in the most peripheral bin
and to the number of binary collisions ($R_{cp}$).
Although, within the large statistical uncertainty, 
they appear to be compatible with a linear scaling with the number of binary collisions, 
the low statistics preclude drawing any 
conclusion.

\section{Conclusion}

The first results on $J/\psi$ and $Z \to \mu^{+} \mu^{-}$ relative yields 
measured in lead-lead collisions obtained with the ATLAS detector at the LHC,
have been presented. 
In a sample of events with oppositely charged muon pairs with 
a transverse momentum above 3 GeV and with $|\eta| < 2.5$, a centrality 
dependent suppression
is observed in the normalized $J/\psi$ yield.
The relative yields of the 38 observed $Z$ candidates as a 
function of centrality are also presented,
although no 
conclusion can be inferred about their scaling with the number
of binary collisions.
\\

\section*{Acknowledgements}

\input{Acknowledgement-29Nov10-collisions.tex}








\include{Biblio}
\clearpage
\label{app:ATLASColl}
\begin{center}
\input{atlas_authlist}

\end{center}

\end{document}

%% file: Acknowledgement-29Nov10-collisions.tex
We  thank CERN for the efficient commissioning and operation of the LHC during this initial heavy ion data taking period as well as the support staff from our institutions without whom ATLAS could not be operated efficiently.

We acknowledge the support of ANPCyT, Argentina; YerPhI, Armenia; ARC, Australia; BMWF, Austria; ANAS, Azerbaijan; SSTC, Belarus; CNPq and FAPESP, Brazil; NSERC, NRC and CFI, Canada; CERN; CONICYT, Chile; CAS, MOST and NSFC, China; COLCIENCIAS, Colombia; MSMT CR, MPO CR and VSC CR, Czech Republic; DNRF, DNSRC and Lundbeck Foundation, Denmark; ARTEMIS, European Union; IN2P3-CNRS, CEA-DSM/IRFU, France; GNAS, Georgia; BMBF, DFG, HGF, MPG and AvH Foundation, Germany; GSRT, Greece; ISF, MINERVA, GIF, DIP and Benoziyo Center, Israel; INFN, Italy; MEXT and JSPS, Japan; CNRST, Morocco; FOM and NWO, Netherlands; RCN, Norway;  MNiSW, Poland; GRICES and FCT, Portugal;  MERYS (MECTS), Romania;  MES of Russia and ROSATOM, Russian Federation; JINR; MSTD, Serbia; MSSR, Slovakia; ARRS and MVZT, Slovenia; DST/NRF, South Africa; MICINN, Spain; SRC and Wallenberg Foundation, Sweden; SER,  SNSF and Cantons of Bern and Geneva, Switzerland;  NSC, Taiwan; TAEK, Turkey; STFC, the Royal Society and Leverhulme Trust, United Kingdom; DOE and NSF, United States of America.  

The crucial computing support from all WLCG partners is acknowledged gratefully, in particular from CERN and the ATLAS Tier-1 facilities at TRIUMF (Canada), NDGF (Denmark, Norway, Sweden), CC-IN2P3 (France), KIT/GridKA (Germany), INFN-CNAF (Italy), NL-T1 (Netherlands), PIC (Spain), ASGC (Taiwan), RAL (UK) and BNL (USA) and in the Tier-2 facilities worldwide.

%% file: atlas_authlist.tex
\begin{flushleft}
{\Large The ATLAS Collaboration}

\bigskip

G.~Aad$^{\rm 48}$,
B.~Abbott$^{\rm 111}$,
J.~Abdallah$^{\rm 11}$,
A.A.~Abdelalim$^{\rm 49}$,
A.~Abdesselam$^{\rm 118}$,
O.~Abdinov$^{\rm 10}$,
B.~Abi$^{\rm 112}$,
M.~Abolins$^{\rm 88}$,
H.~Abramowicz$^{\rm 153}$,
H.~Abreu$^{\rm 115}$
% ...
\bigskip

% ...
$^{10}$ Institute of Physics, Azerbaijan Academy of Sciences, H. Javid Avenue 33, AZ 143 Baku, Azerbaijan\\
$^{11}$ Institut de F\'isica d'Altes Energies, IFAE, Edifici Cn, Universitat Aut\`onoma  de Barcelona,  ES - 08193 Bellaterra (Barcelona), Spain\\
% ...
$^{48}$ Albert-Ludwigs-Universit\"{a}t, Fakult\"{a}t f\"{u}r Mathematik und Physik, Hermann-Herder Str. 3, D - 79104 Freiburg i.Br., Germany\\
$^{49}$ Universit\'e de Gen\`eve, Section de Physique, 24 rue Ernest Ansermet, CH - 1211 Geneve 4, Switzerland\\
% ...
$^{88}$ Michigan State University, Department of Physics and Astronomy, High Energy Physics Group, East Lansing, MI 48824-2320, United States of America\\
% ...
$^{111}$ University of Oklahoma, Homer L. Dodge Department of Physics and Astronomy, 440 West Brooks, Room 100, Norman, OK 73019-0225, United States of America\\
$^{112}$ Oklahoma State University, Department of Physics, 145 Physical Sciences Building, Stillwater, OK 74078-3072, United States of America\\
% ...
$^{115}$ LAL, Univ. Paris-Sud, IN2P3/CNRS, Orsay, France\\
% ...
$^{118}$ Oxford University, Department of Physics, Denys Wilkinson Building, Keble Road, Oxford OX1 3RH, United Kingdom\\
% ...
$^{153}$ Tel Aviv University, Raymond and Beverly Sackler School of Physics and Astronomy, Ramat Aviv, IL - Tel Aviv 69978, Israel\\
% ...
\end{flushleft}